\documentclass[aps,prl,twocolumn]{revtex4}
\usepackage{graphicx}

\begin{document}

\title{Anomalous Pressure Dependence of Kadowaki-Woods Ratio and Crystal Field Effects in Mixed-Valence YbInCu$_{4}$}
\author{Tuson Park$^{1}$}
\author{V. A. Sidorov$^{1}$}
\altaffiliation{On leave from Vereshchagin Institute of High Pressure Physics, 142190 Troitsk, Russia}
\author{J. L. Sarrao$^{1}$}
\author{J. D. Thompson$^{1}$}
\affiliation{$^{1}$Los Alamos National Laboratory, Los Alamos, New Mexico 87545, USA}
\date{\today}

\begin{abstract}
The mixed-valence (MV) compound YbInCu$_{4}$ was investigated by electrical resistivity and ac specific heat at low temperatures and high pressures. At atmospheric pressure, its Kadowaki-Woods (KW) ratio, $A/\gamma ^{2}$, is 16 times smaller than the universal value $R_{\text{KW}}$($=1.0\times 10^{-5} \mu \Omega \cdot$cm$\cdot$mol$^{2}\cdot$K$^{2}\cdot$mJ$^{-2}$), but sharply increases to $16.5R_{\text{KW}}$ at 27~kbar. The pressure-induced change in the KW ratio and deviation from $R_{\text{KW}}$ are analyzed in terms of the change in f-orbital degeneracy $N$ and carrier density $n$. This analysis is further supported by a dramatic change in residual resistivity $\rho_{0}$ near 25~kbar, where $\rho_{0}$ jumps by a factor of 7.
\end{abstract}

\maketitle
At low temperatures, strongly correlated metals exhibit an enhanced electronic specific heat that is a measure of the density of electronic states at the Fermi energy $N(E_{\text{F}})$, which results from hybridization of nearly-localized and ligand electrons. Scattering of electrons within this strongly correlated Fermi-liquid state is characterized by an electrical resistivity that increases from $T=0$ as temperature squared, that is, $\rho(T) = \rho(T=0) +AT^2$. Kadowaki and Woods \cite{kadowaki86} showed that the ratio of the $T^2 -$resistivity coefficient to specific heat Sommerfeld coefficient $\gamma$ assumed a common value $A/\gamma ^{2} = R_{\text{KW}}=1.0\times 10^{-5} \mu \Omega \cdot$cm$\cdot$mol$^{2}\cdot$K$^{2}\cdot$mJ$^{-2}$ for a diverse set of strongly correlated materials with $\gamma$ values ranging by two orders of magnitude. YbInCu$_4$, which has a low-temperature $\gamma$ well within the range of these other materials, is an interesting exception to this common ratio. Such departures from the usual value of $R_{\text{KW}}$ challenge Fermi-liquid predictions for strongly correlated systems \cite{kontani04}.

At temperatures above $T_{\text{v}}=42$~K, YbInCu$_4$ is characterized by a Curie-Weiss temperature dependence in magnetic susceptibility and a semi-metallic behavior in transport measurements due to its localized 4f electrons \cite{felner87,nakamura90,nakamura94}. At $T_{\text{v}}$, there is a first-order isostructural valence transition accompanied by a volume increase of +0.5\%. In contrast to the trivalent configuration of Yb's 4f electrons above $T_{\text{v}}$, the volume expansion induces strong charge hybridization, a change in Yb's valence from 3+ at $T> T_{\text{v}}$ to a mixed-valence value of 2.9+ below $T_{\text{v}}$~\cite{felner87}, and an enhanced Pauli-like susceptibility corresponding to a modestly large $\gamma=50$~mJ/mol$\cdot$K$^2$~\cite{sarrao98}. The $T^2 -$coefficient of resistivity at $T<<T_{\text{v}}$, however, is much smaller than anticipated by the universal value $R_{\text{KW}}$. Decreasing the cell volume of YbInCu$_4$ by applied pressure stabilizes the high-temperature 4f$^{13}$ configuration to progressively lower temperatures and eventually drives $T_{\text{v}}$ to $T=0$ near 25~kbar~\cite{mito03}, thereby enabling systematic exploration of the origin of the anomalous Kadowaki-Woods ratio that has been observed in other mixed-valence Yb and Ce-based compounds~\cite{tsujii03}. In this Letter, we report a pressure-induced KW-ratio change of YbInCu$_4$ from 0.063$R_{\text{KW}}$ at $P=0$~kbar to $16.5R_{\text{KW}}$ at $P=27$~kbar and show that the pressure dependence of the KW ratio and deviation from $R_{\text{KW}}$ are due to the change in f-orbital degeneracy $N$ and carrier density $n$. This conclusion is further supported by a dramatic change in residual reisitity $\rho_{0}$ near 25~kbar, where $\rho_{0}$ jumps by a factor of 7.

Single crystals of YbInCu$_{4}$ were obtained from an In-Cu flux method as described elsewhere~\cite{sarrao96}. Hydrostatic pressure was achieved by using a hybrid Be-Cu/NiCrAl clamp-type pressure cell with silicone fluid as a transmitting medium. The absolute value of pressure at low temperatures was obtained by measuring the superconducting transition temperature of a tin manometer. The resistivity $\rho$ of YbInCu$_{4}$ was measured by a standard four-point method with a LR-700 ac resistance bridge (Linear Research). At ambient pressure, $\rho$ has a very narrow mixed-valence transition region, $\Delta T \leq 1$~K, and a large resistivity ratio, $\rho (T_{\text{v+}}) / \rho (T_{\text{v-}}) \approx 13$, suggesting that the analyzed crystal is of high quality~\cite{sarrao96}. Here, $\rho (T_{\text{v+}})$ is the resistivity at temperatures just above $T_{\text{v}}$ and $\rho (T_{\text{v-}})$, below $T_{\text{v}}$.

In Fig.~1(a), we show the resistivity of YbInCu$_{4}$ as a function of temperature at several pressures on a log-log scale. The valence transition temperature $T_{\text{v}}$ is rapidly suppressed by applying pressure, $dT_{\text{v}}/dP \approx -1.7$~K/kbar, as reported in transport and thermodynamic measurements~\cite{matsumoto92,sarrao98a,hedo03}, and goes to zero near 25~kbar. At $P=27$~kbar, there is no clear drop in $\rho$ that can be attributed to a valence transition. The temperature derivative of the resistivity, surprisingly, reveals a scaling behavior among different pressure data (Fig.~1b). Three distinct regimes in $d\rho /dT$ are identified by a sharp change in the slope at $T_{\text{v}}$ and 32~K: regime~I for $T > 32$~K, regime~II for $T_{\text{v}} < T < 32$~K, and regime~III for $T < T_{\text{v}}$. 
\begin{figure}[tbp]
\centering  \includegraphics[width=8cm,clip]{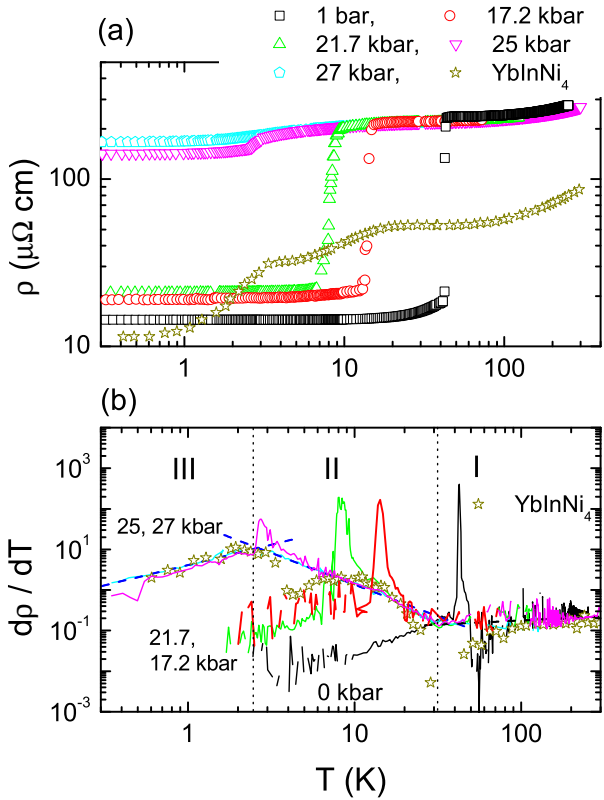}
\caption{(color online) (a): Resistivity vs. temperature on a log-log scale of YbInCu$_{4}$ at different pressures. For comparison, $\rho$ of YbInNi$_{4}$ at 1~bar is shown as open stars. (b): Temperature derivative of the resistivity $d\rho /dT$. For clarity, only a few representative data sets are shown. The dashed lines in regime II and III represent $1/\sqrt{T}$ and $T^2$ behaviors, respectively.}
\label{figure1}
\end{figure}

In the high temperature phase (regime~I), the electrical resistivity can be accounted for by phonon $\rho_{\text{ph}}$ and spin-dependent scattering processes $\rho_{\text{mag}}$: $\rho = \rho_{0}+\rho_{\text{ph}}(T)+\rho_{\text{mag}}(T)=\rho_{0}+b~T-c~lnT$. The good fit of the above model and a shallow resistivity minimum observed at 55~K and ambient pressure (not shown here) signify the importance of Kondo interactions in this compound. The resistivity slope is almost independent of temperature. 

For all pressures $P\geq 17.2$~kbar, where $T_v < 32$~K, the resistivity exhibits a significant increase in slope below 32~K. This temperature is of the same order as the crystalline electric field splitting 3.2~meV found in the high temperature trivalent state by inelastic neutron scattering measurements in YbInCu$_4$~\cite{severing90}. From neutron scattering measurements~\cite{severing90}, the stable trivalent compound YbInNi$_4$ exhibits essentially identical crystal field splitting. The de-excitation of the excited crystal field level causes a broad shoulder near $20-30$~K in the resistivity of YbInNi$_4$ (Fig.~1a) which gives rise to the sharp increase in $d\rho /dT$ below 32~K (Fig.~1b). The sharp drop in $\rho$ below 3.2~K in YbInNi$_4$ is due to ferromagnetism. The scaling behavior observed in Fig.~1b in region~II is thus a property of the high temperature $(T>T_v)$ trivalent state which is already present in YbInNi$_4$. The existence of this scaling behavior also implies that the crystal field splitting does not change significantly with pressure. 

Specific heat measurements of YbInCu$_4$ also support the conclusion that 32~K is the crystal field splitting~\cite{sidorov05}. Figure~2 shows representative specific heat at 27.9~kbar, where a sharp specific heat anomaly due to ferromagnetism occurs at 2.5~K \cite{mito03} and a broad maximum near 10~K is attributable to the combination of Kondo (dotted line)~\cite{rajan83} and crystal field (dash-dotted line) effects. The best fit to the broad maximum was obtained with Kondo temperature $T_K =18$~K and crystal field splitting $\Delta =28$~K with a quartet ground state. A Sommerfeld coefficient $\gamma$(=110$\pm$50 mJ/mol$\cdot$K$^2$) is obtained from a least squares fit to data below $T_c$ (solid line).
\begin{figure}[tbp]
\centering  \includegraphics[width=8cm,clip]{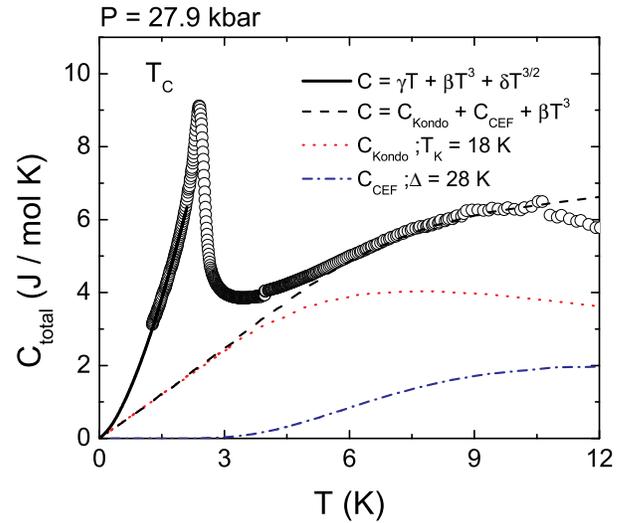}
\caption{(color online) Specific heat versus $T$ at 27.9~kbar \cite{sidorov05}. The broad maximum near 10~K was interpreted by Kondo (dotted line) and crystal-field (dash-dotted line) effects with a quartet ground state and a doublet excited state. Sommerfeld coefficient $\gamma$ is obtained by fitting $C_{total}=\gamma T+\beta T^3 +\delta T^{3/2}$ below $T_c$ (=2.5~K), where the third term is due to magnon contributions.}
\label{figure2}
\end{figure}

In the low temperature phase below $T_{\text{v}}$ (regime~III), the temperature dependence of $\rho$ shows typical Fermi-liquid behavior: $\rho =\rho _{0}+AT^{2}$. In Fig.~3(a) we show scaling of the normalized resistivity $\Delta \rho /AT_{0}^{2}$ versus $(T/T_{0})^{2}$. Here, $\Delta \rho = \rho (T) - \rho _{0}$ and $T_{0}$ is the highest temperature where the $T^{2}$ dependence is obeyed. The scaling temperature $T_{0}$ is typically associated with a coherence temperature $T_{\text{co}}$, which is proportional to the Kondo temperature, i.e., $T_{\text{co}}\propto T_{\text{K}}$~\cite{andres75,assaad04}. It is interesting to note that the presence of the mixed-valence transition changes the resistivity behavior above $T_{0}$: the normalized resistivity deflects upward for $P\leq 25$~kbar, while it deviates downward for $P>25$~kbar. YbInNi$_{4}$ whose ground state is ferromagnetic follows the 27~kbar data of YbInCu$_{4}$.
\begin{figure}[tbp]
\centering  \includegraphics[width=8cm,clip]{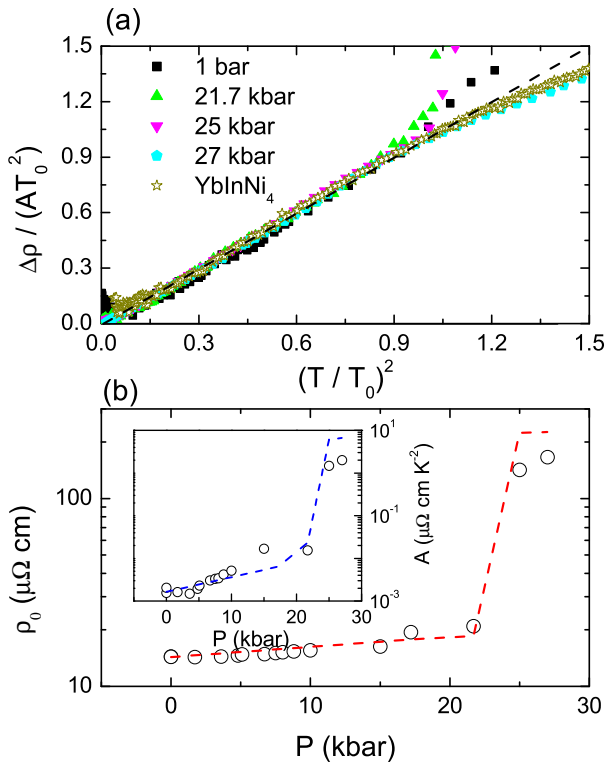}
\caption{(color online) (a) Normalized resistivity of YbInCu$_{4}$ versus $(T/T_{0})^{2}$ at low temperatures and various pressures, where $T_{0}$ is the highest temperature over which $\rho$ obeys a $T^{2}$ dependence, $\Delta \rho = \rho (T) - \rho (T=0)$, and $A$ is the $T^{2}$ coefficient. The dashed line is a guide to the eye. (b) Pressure dependence of the residual resistivity on a semi-log scale. The dashed line is a fit to Eq.~(1). Inset: pressure dependence of $A$ on a semi-log scale. The dashed line is a fit to Eq.~(2)(see text).}
\label{figure3}
\end{figure}

The pressure dependence of the residual resistivity $\rho _{0}(P)$ obtained from the $T^{2}$ scaling is displayed in Fig.~3(b). For $P<25$~kbar, $\rho_{0}$ increases from 14.4 at $P=0$ to 20.9~$\mu \Omega \cdot$cm at $P=21.7$~kbar. The increase in $\rho _{0}$ over this pressure interval is comparable to that of other heavy fermion compounds and suggests that the residual scattering is not simply due to potential scattering from impurities or lattice disorder, but involves the Yb-4f electrons~\cite{lawrence85}. When lattice periodicity is disrupted by disorder in Yb or In sites of YbInCu$_{4}$~\cite{sarrao96}, the scattering phase shift of the Yb-4f electrons fluctuates, resulting in a finite resistivity at $T=0$~K~\cite{newns81,terzieff98}:
\begin{equation}
\rho _{0}=\frac{G}{n^{1/3}}~ sin^{2}(\pi n_{\text{f}}/N),
\end{equation}
where $N$ is the effective f-orbital degeneracy and $G$ is a prefactor that takes into account the pressure effects of the hybridization strength. Pressure dependence of the effective 4f hole occupation number $n_{\text{f}}$ of YbInCu$_{4}$ was obtained from reference~\cite{mushnikov04} up to 16~kbar. For $P>16$~kbar, $n_{\text{f}}$ was extrapolated with an assumption that $n_{\text{f}}$ at 25~kbar is equal to that $(=0.94)$ in the high temperature phase at $P=0$~kbar~\cite{mushnikov04}. For $T<T_{\text{v}}$, the effective degeneracy $N$ is 8 $(=2j+1)$, as expected from the strongly hybridized MV state and from the absence of CEF excitations in neutron scattering~\cite{severing90}, and where $j$ is the total angular momentum $(=7/2)$. Even though the carrier density $n$ is a function of pressure, for simplicity, we used 2.2 per formula unit below 25~kbar and 0.07 above that pressure~\cite{cornelius97}. The best fit was obtained with $G(P)\propto (1+0.005P)$ (dashed line in Fig.3~b), which is similar to that of YbInNi$_{4}$~\cite{g-factor}. The positive slope in $\rho _{0}$ of the Yb-based compounds, $d\rho _{0}/dP >0$, reflects the 4f-hybridization effects, which make the slope negative in Ce-based compounds, electron analogs of Yb-based compounds.

At $P=25$~kbar, there is a pronounced jump in $\rho _{0}$ by a factor of 7. Once the characteristic energy scale for spin fluctuations decreases with applied pressure below the first excited CEF level, the effective degeneracy will decrease and, consequently, the residual resistivity increases sharply. When the effective degeneracy changes from 8 to 4, the sharp increase is well explained by Eq.~(1) (dashed line in Fig.~3b). Further support for a change in degeneracy comes from the pressure dependence of the Kondo temperature $T_{\text{K}}$ and crystal-field scale $T_{\text{CEF}}$ (see Fig.~4a). $T_{\text{K}}$ decreases with pressure, while $T_{\text{CEF}}$ is essentially independent of pressure. $T_{\text{K}}$, indeed, becomes smaller than $T_{\text{CEF}}$ near 25~kbar, indicating that crystal-fields which lift the full $j-$manifold degeneracy, are important for $P\geq 25$~kbar. Values of $T_{\text{K}}$ plotted in Fig.~4(a) were obtained from reference~\cite{sarrao98a,mushnikov04} for $P<25$~kbar and specific heat measurements for $P\geq 25$~kbar~\cite{sidorov05}.

A change in ground state degeneracy is also manifested in the pressure dependence of the $T^{2}$ coefficient $A$. The gradual increase in $A$ with pressure turns into a jump at $P=25$~kbar. In the inset of Fig.~3(b), we compare experimental results to predictions for a lattice Anderson model \cite{coleman87}:
\begin{equation}
A = \frac{\kappa}{N^{2}}~ \frac{n^{-1/3}}{T_{\text{K}}^{2}}\hspace{0.5cm},
\end{equation}
where $\kappa$ is constant and $n$ is carrier density. Best results were obtained when the degeneracy $N$ changes from 8 to 4 near 25~kbar, which is consistent with the quartet ground state claimed both from inelastic neutron scattering~\cite{severing90} and from thermodynamic measurements \cite{sarrao98}.
\begin{figure}[tbp]
\centering  \includegraphics[width=8cm,clip]{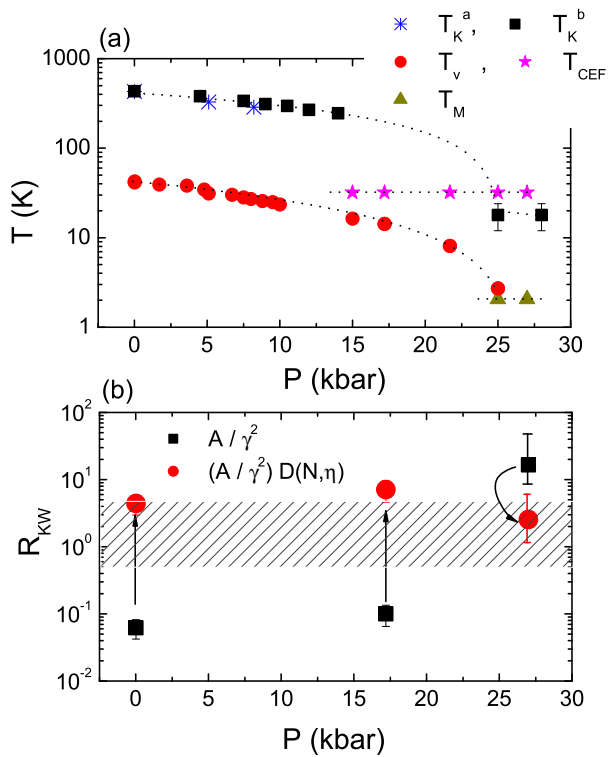}
\caption{(color online) (a) $T-P$ phase diagram on a semi-log scale. $T_{\text{M}}$ is a magnetic ordering temperature from $\chi_{\text{ac}}$ (not shown) and $T_{\text{v}}$ is the valence-transition temperature. The Kondo temperature $T_{\text{K}}^{a}$ is from reference~\cite{sarrao98a} and $T_{\text{K}}^{b}$ from reference~\cite{mushnikov04}. For $P>25$~kbar, $T_K$ is obtained from specific heat measurements \cite{sidorov05}. The dotted line that connects $T_{\text{K}}$'s is a guide to the eyes \cite{guideline}. (b) The pressure dependence of the KW ratio $A/\gamma ^2$ is shown in squares and the renormalized ratio $(A/\gamma ^2)D(N,\eta)$, in circles. The y-axis unit is $R_{KW}(=1.0\times 10^{-5} \mu \Omega \cdot$cm$\cdot$mol$^{2}\cdot$K$^{2}\cdot$mJ$^{-2})$. The hatched band is from 0.5 to 5~$R_{\text{KW}}$ within which most of heavy fermions are located.}
\label{figure4}
\end{figure} 

Recently, a correlation between the KW ratio and a f-orbital degeneracy $N$ of heavy fermion compounds was theoretically explored by Kontani \cite{kontani04} who predicted:
\begin{equation}
 \frac{A}{\gamma ^{2}}\approx D^{-1}(N,\eta)R_{\text{KW}}=\frac{2R_{\text{KW}}}{N(N-1)} \frac{\alpha}{\eta^{4/3}},
\end{equation}
where the lattice spacing $a=\alpha a_0$ with $a_0=1\times 10^{-8}$cm and carrier density $n=\eta / a^3$. When the degeneracy $N$ changes, it not only changes the electronic density of states but also the scattering rates, renormalizing both quantities $A$ and $\gamma$. Figure~4b shows the KW ratio before (squares) and after (circles) taking into account $D(N,\eta)$. The anomalously small measured value of $A/\gamma ^{2}$ ($~0.063R_{\text{KW}}$) at $P=0$~kbar increases to $16.5R_{\text{KW}}$ at $P = 27$~kbar, where the effective degeneracy changes from an octet to a quartet. When both the degeneracy factor and carrier density are considered (circles), $(A/\gamma ^{2})D(N,\eta)$, the renormalized ratios at all pressures converge to the characteristic value found in most heavy fermion compounds \cite{kadowaki86}, delineating the relationship between the effective degeneracy and the small value of the KW ratio.

In summary, we have reported an anomalous pressure dependence of the KW ratio of YbInCu$_{4}$. At $P=0$~kbar, the Kadowaki-Woods (KW) ratio, $A/\gamma ^{2}$, is 16 times smaller than the universal value $(R_{\text{KW}})$, but sharply increases to $16.5R_{\text{KW}}$ at 27~kbar. The pressure-induced change in the KW ratio and deviation from $R_{\text{KW}}$ are analyzed in terms of the change in $f$-orbital degeneracy $N$ and carrier density $n$, which was further supported by a pronounced change in the residual resistivity at 25~kbar. We claim that there no longer should be a distinction between `anomalous' and `normal' values of the KW ratio; rather, both the smaller and more commonly observed values are a direct consequence of carrier density and magnetic degeneracy of $f$-ions.

Work at Los Alamos was performed under the auspices of the U.S. Department of Energy.  V.A.S. acknowledges the support of Russian Foundation for Basic Research (Grant No. 03-02-17119) and Program ``Physics and Mechanics of Strongly Compressed Matter of Presidium of Russian Academy of Sciences''. We acknowledge benefits from discussion with J. M. Lawrence and I. Vekhter.


\begin{thebibliography}{23}
\bibitem{guideline} The shape of the guide line is consistent with reference [19] which showed that the pressure dependence of the Kondo temperature is similar to that of a characteristic mixed-valence temperature.
\end{thebibliography}

\end{document}